\documentclass{article}

\usepackage[utf8]{inputenc}

\usepackage[english]{babel}
\usepackage{subeqnarray}
\usepackage{url}

\usepackage{graphicx,psfrag}
\usepackage{fixltx2e,amsmath}
\usepackage{xcolor}
\usepackage{amssymb}
\usepackage[final]{pdfpages}
\usepackage{bbold}
\usepackage{dsfont}
\usepackage{pgfplots}
\usepackage{subfig}
\usepackage{algorithm}
\usepackage{algorithmic,placeins}
\usepackage[font=small,labelfont=bf]{caption}

\usepackage[font={footnotesize}]{caption}

\usepackage[utf8]{inputenc}
\usepackage[T1]{fontenc}
\usepackage{xcolor,amsmath,amssymb}
\usepackage{geometry}
\usepackage{graphicx}
\usepackage[sort&compress,numbers]{natbib}
\usepackage{enumerate}
\usepackage{wasysym}
\usepackage[textsize=footnotesize]{todonotes}

\usepackage{hyperref}
\hypersetup{
    colorlinks=true,
    linkcolor=red,
    filecolor=magenta,      
    urlcolor=cyan,
}

\usepackage{pgfplots}

\definecolor{darkblue}{rgb}{0,0,0.6}
\definecolor{darkred}{rgb}{0.6,0,0}

\usepackage{hyperref}

\def\Re{{\rm Re}\,}\def\Im{{\rm Im}\,}

\def\to{\rightarrow}

\newcommand{\beq}{\begin{equation}} \newcommand{\eeq}{\end{equation}}

\newcommand\be{\begin{equation}}
\newcommand\bea{\begin{eqnarray} \nonumber }
\newcommand\ee{\end{equation}}
\newcommand\eea{\end{eqnarray}}

\usepackage{authblk}

\usepackage[bitstream-charter]{mathdesign}

\usetikzlibrary{pgfplots.groupplots}

\DeclareMathOperator{\tr}{Tr}

\graphicspath{{figures/}} 

\title{On a Generalisation of the Mar\v cenko-Pastur Problem}
\author{Jean-Philippe Bouchaud}
\affil{Capital Fund Management \& Acad\'emie des Sciences, Paris} 
\author{Marc Potters}
\affil{Capital Fund Management, Paris} 
\date{September 2020}

\begin{document}

\maketitle
\begin{abstract}
We study the spectrum of generalized Wishart matrices, defined as $\mathbf{F}=( \mathbf{X} \mathbf{Y}^\top + \mathbf{Y} \mathbf{X}^\top)/2T$, where $\mathbf{X}$ and $\mathbf{Y}$ are $N \times T$ matrices with zero mean, unit variance IID entries and such that $\mathbb{E}[\mathbf{X}_{it} \mathbf{Y}_{jt}]=c \delta_{i,j}$. The limit $c=1$ corresponds to the Mar\v cenko-Pastur problem. For a general $c$, we show that the Stieltjes transform of $\mathbf{F}$ is the solution of a cubic equation. In the limit $c=0$, $T \gg N$ the density of eigenvalues converges to the Wigner semi-circle.   
\end{abstract}

The celebrated Mar\v cenko-Pastur problem concerns the eigenvalue spectrum of random covariance matrices in the large dimension limit. More precisely, consider an N-dimensional time series, $x_i^t$, where $i=1, \cdots, N$ and $t=1, \cdots, T$. Suppose that the $x$'s are IID random variables, of zero mean and unit variance. Then the empirical (or sample) covariance matrix is defined as 
\begin{equation}
\mathbf{E} := \frac{1}{T} \sum_{t=1}^T x_i^t x_j^t = \frac{1}{T} \mathbf{X} \mathbf{X}^\top ,
\end{equation}
where $\mathbf{X}$ is the $N \times T$ matrix defined by $\mathbf{X}_{it}=x_i^t$. $\mathbf{E}$ is called a (white) Wishart matrix. 

Since our assumption is that the $x$'s are IID, the ``true'' covariance matrix is simply the identity matrix, which is the result obtained for $\mathbf{E}$ in the limit $T \to \infty$, for a fixed value of $N$. But there is another limit, which is very relevant in many applications, where $T$ and $N$ are both large; more precisely, where $N,T \to \infty$ with a fixed ratio $q=N/T$. 

What is the spectrum of $\mathbf{E}$ in that regime? The answer was provided by Mar\v cenko and Pastur in 1967 \cite{MP}, and is a classic result in Random Matrix Theory. When $q < 1$, the result for the density of eigenvalues $\rho(\lambda)$ is:
\begin{equation}
\rho(\lambda) = \frac{\sqrt{(\lambda_+-\lambda)(\lambda-\lambda_-)}}{2 \pi q \lambda}, \qquad \lambda_- \leq \lambda \leq \lambda_+,
\end{equation}
with $\lambda_\pm = (1\pm\sqrt{q})^2$. As expected $\lambda_+=\lambda_-=1$ when $q \to 0$, i.e. when $T \gg N$, $\mathbf{E}$ becomes the identity matrix. 

The problem we want to consider in this note is defined by the following symmetric cross-correlation matrix:\footnote{The singular value spectrum of the {\it un-symmetrized} matrix $\mathbf{X} \mathbf{Y}^\top/T$ was considered for $c=0$ in \cite{Wachter,Micelli}.}
\begin{equation}
\mathbf{F} := \frac{1}{2T} \left( \mathbf{X} \mathbf{Y}^\top + \mathbf{Y} \mathbf{X}^\top \right)
\end{equation}
where $\mathbf{X}$ and $\mathbf{Y}$ are two $N \times T$ rectangular matrices of unit variance IID random variables, such that $\mathbb{E}[x_i^t y_j^t]=c \delta_{i,j}$, where $c \in [-1,1]$ is the correlation coefficient between $x$'s and $y$'s. Clearly, when $c=1$, $\mathbf{X}=\mathbf{Y}$ and one recovers the Mar\v cenko-Pastur problem. The aim of this work is to determine the spectrum of $\mathbf{F}$.

The most efficient way to approach this problem is by using the tools of free random matrix theory (see e.g. \cite{Verdu,CUP}). In the large dimension limit, $\mathbf{F}$ can be seen as the free addition of T two-dimensional projectors:
\begin{equation}
\mathbf{F} = \sum_t \mathbf{P}_t, \qquad (\mathbf{P}_t)_{ij}:=\frac{1}{2T} \left(x^t_i y^t_j + y^t_i x^t_j \right).
\end{equation}
In the large $N$ limit, the spectrum of $\mathbf{P}_t$ becomes independent of $t$ and is composed of $N-2$ zero eigenvalues and two non-zero eigenvalues, equal to $q(c \pm 1)/2$. The corresponding Stieltjes function is thus:
\begin{equation}
G(z) = \frac{N-2}{N}\, \frac{1}{z} + \frac{1}{N} \left(\frac{1}{z - q(c+1)/2} + \frac{1}{z - q(c-1)/2} \right),
\end{equation}
from which we deduce the functional inverse $\mathfrak{z}(u)$ defined as $\mathfrak{z}(G(z))=z$. To leading order in $1/N$ one finds:
\begin{equation}
\mathfrak{z}(u) = \frac{1}{u} + \frac{1}{N} \frac{4cq + 2q^2 u (1-c^2)}{(2 - cqu)^2 - q^2 u^2} + O(N^{-2}).
\end{equation}
Free matrix addition means that the $R$-transform, defined as $R(u)=\mathfrak{z}(u)-u^{-1}$ is additive. Hence the R-transform of $\mathbf{F}$ is given by
\begin{equation}
R_{\mathbf{F}}(u) =  \frac{4c + 2q u (1-c^2)}{(2 - cqu)^2 - q^2 u^2}, \qquad N \to \infty.
\end{equation}
\begin{figure}
\centering
\includegraphics{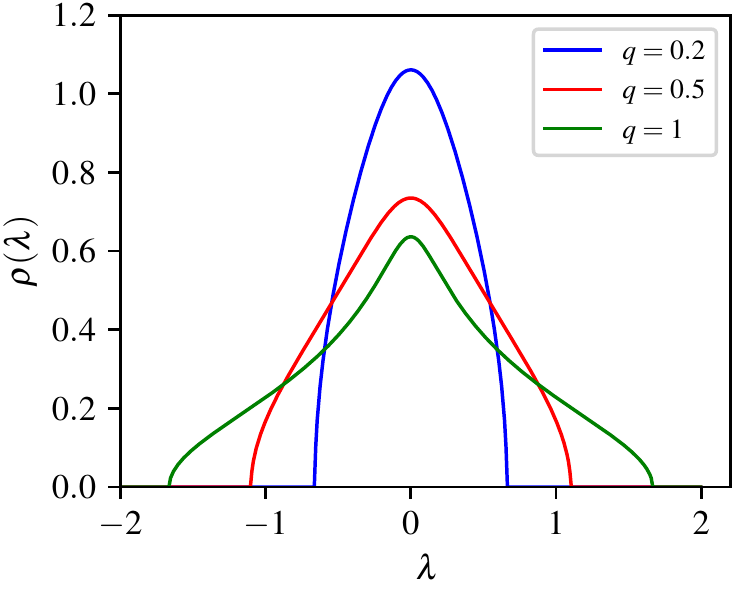}
\includegraphics{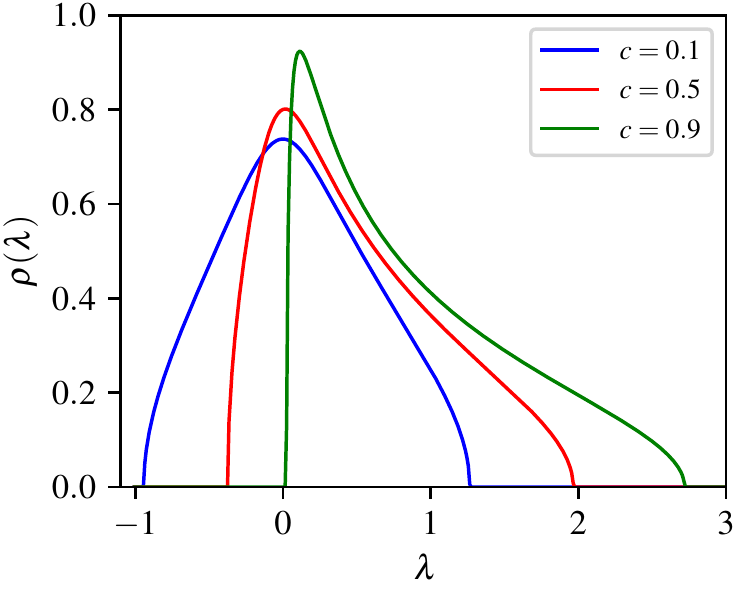}
\caption{Analytical densities of eigenvalues, $\rho(\lambda)$, for $c=0$, different $q$'s (left) and $q=0.5$, different $c$'s (right). The density for $c=0, q=1$ (green curve left) is known as the ``Tetilla law'' as its shape is similar to that of a Galician cheese.}\label{fig1}
\end{figure}
\begin{figure}
\centering
\includegraphics{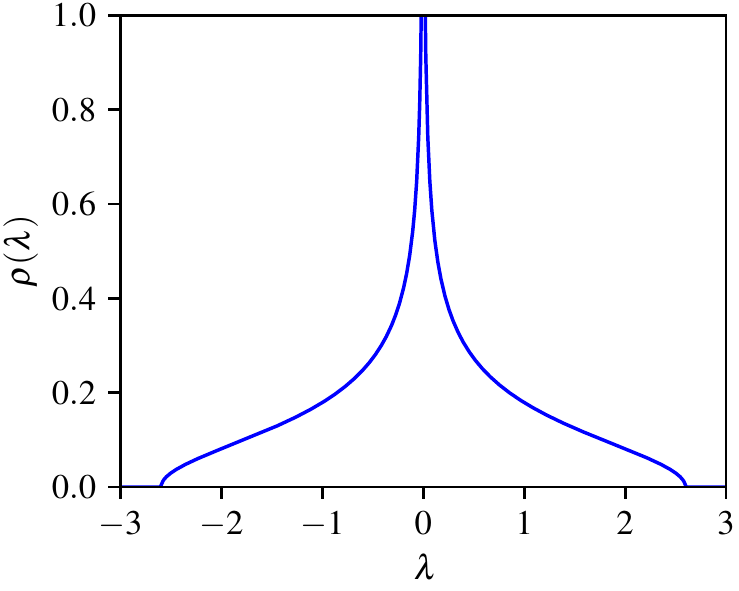}
\includegraphics{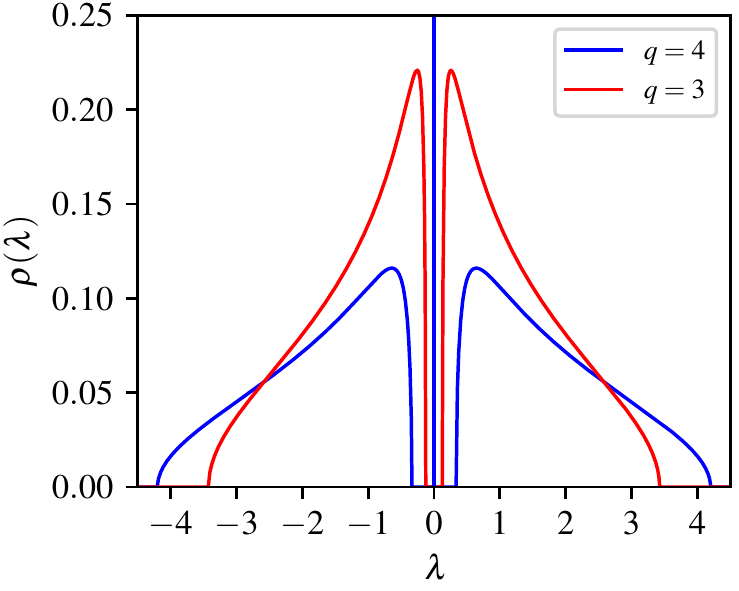}
\caption{Analytical densities of eigenvalues, $\rho(\lambda)$, for the special case $c=0,q=2$ (left) and for $c=0, q > 2$ (right). Note the $|\lambda|^{-1/3}$ singularity for $q=2$, beyond which a gap opens with two inner edges, and a Dirac mass at $\lambda=0$.}\label{fig2}
\end{figure}

From $R_{\mathbf{F}}(u)$ one backtracks to get $\mathfrak{z}_{\mathbf{F}}(u)$ and finally $G_{\mathbf{F}}(z)$, the Stieltjes transform of $\mathbf{F}$ that contains all the information about the density of eigenvalues of $\mathbf{F}$. Finally, $G_{\mathbf{F}}(z)$ is given by the appropriate solution of the following cubic equation:
\begin{equation}\label{cubic}
\frac{q^2 (1-c^2)}{4} z G_{\mathbf{F}}^3 + \left[\frac{q(2-q) (1-c^2)}{4} + cq z\right] G_{\mathbf{F}}^2 - (z + c(q-1)) G_{\mathbf{F}} + 1 = 0.
\end{equation}
This is the main result of the present paper. One can  check that when $c= 1$ one recovers the Stieltjes transform of a Wishart matrix, solution of the following quadratic equation:
\begin{equation}
q z G_W^2 - (z + q-1) G_W + 1 = 0,
\end{equation}
from which the Mar\v cenko-Pastur result immediately follows. In Fig. \ref{fig1} we show the spectrum of $\mathbf{F}$ for some representative values of $c$ and $q$, obtained as usual from the imaginary value of $G_{\mathbf{F}}$ when $\Im(z) \to 0$. Note that when $c \neq 1$, a fraction of the eigenvalues are negative. In fact, the mean of the density of eigenvalues is equal to $c$. The variance can be read-off the R-transform using $\sigma^2=R'(0)=q(1+c^2)/2$.

The case $c=0$ is particularly interesting. In this case the spectrum is an even function of $\lambda$, see Fig. \ref{fig1}. After a little work one can establish the following results:
\begin{figure}
\centering
\includegraphics{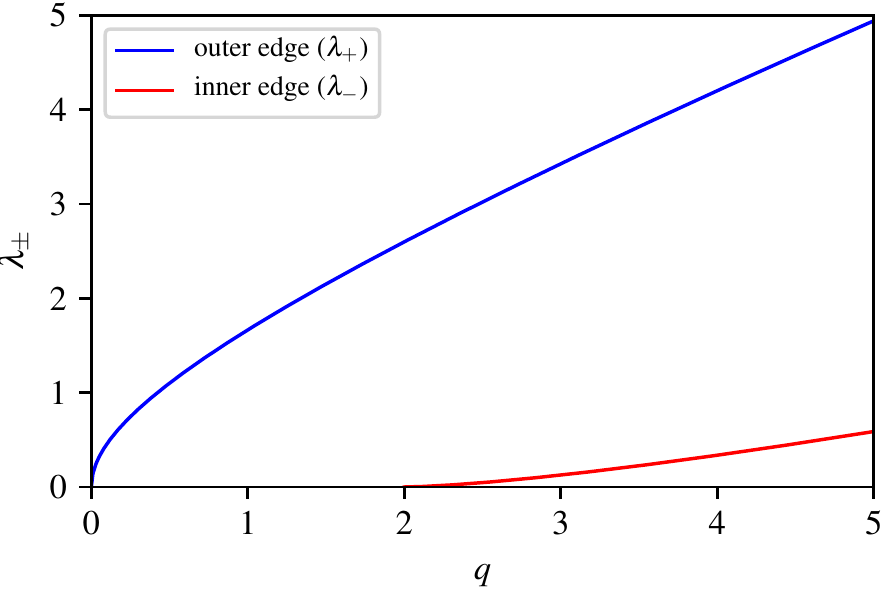}
\caption{Plot of the inner and outer edges as a function of $q$ for $c=0$. Note that the inner edges only exist for $q > 2$.}\label{fig3}
\end{figure}

\begin{enumerate}

    \item The edges of the spectrum, $\lambda_\pm$, are given by:
    \begin{equation}
    \lambda_\pm=\frac{1}{2\sqrt{2}}\sqrt{2q^2 + 10q - 1 \pm \sqrt{64q^3 + 48q^2 +12q+1}},
    \end{equation}
    where $\pm\lambda_+$ are the outer edges and $\pm\lambda_-$ are the inner edges (which only exist for $q>2$). They are plotted in Fig. \ref{fig3}.

    \item In the limit $q \to 0$, the density becomes a Wigner semi-circle of radius $\sqrt{2q}$,
    \begin{equation}
\rho_{q \to 0,c=0}(\lambda) = \frac{\sqrt{2q-\lambda^2}}{\pi q}.
\end{equation}
When $q=0$, one finds that all eigenvalues are zero, as expected since in this case $\mathbf{F} \equiv \mathbf{0}$.\footnote{More generally, when $q=0$, all eigenvalues are equal to $c$.}
\item The case $q=1, c=0$ was studied before in the context of the addition of two non-symmetrized Wigner matrices, see \cite{Nica,tetilla}. The corresponding distribution of eigenvalues was called the ``tetilla'' law \cite{tetilla}.\footnote{It is also the distribution of eigenvalues of $\mathbf{XX}^\top - \mathbf{YY}^\top$, where $\mathbf{X}$ and $\mathbf{Y}$ are independent Wigner matrices.} The explicit form for the density when $|\lambda|\leq\lambda_+=2^{-3/2}\sqrt{11+5\sqrt{5}}$ is given by:
\begin{equation}
    \rho_{q =1,c=0}(\lambda)= \frac{1}{2\sqrt{3}\pi|\lambda|}\left(u_+^{1/3}-u_-^{1/3}\right),\quad
    u_\pm=1+72\lambda^2\pm3\sqrt{12\lambda^2+528\lambda^4-192\lambda^6}.
\end{equation}
Note that with our normalisation it has $\sigma^2=1/2$.
\item For $q=2$, we also have a  closed form expression for the density supported on $|\lambda|\leq\lambda_+$:
\begin{equation}
    \rho_{q =2,c=0}(\lambda)= \frac{1}{2\pi}\left(u^{-1}-u\right),\quad u=\left(\frac{\lambda_+-\sqrt{\lambda_+^2-\lambda^2}}{|\lambda|}\right)^{1/3},\quad\lambda_+=\frac{3\sqrt{3}}{2}.
\end{equation}
This density has a cubic-root singularity at $\lambda=0$ (see Fig. \ref{fig2}):
\begin{equation}
\rho_{q =2,c=0}(\lambda \to 0) \sim \frac{\sqrt{3}}{2\pi |\lambda|^{1/3}}.
\end{equation}
\item For $q > 2$, a Dirac mass appears at $\lambda=0$. This is expected since in that case the $T$ two-dimensional projectors $\mathbf{P}_t$ can no longer span the whole $N$-dimensional space. 
\end{enumerate}

The above construction does not rely on the fact that the matrices $\mathbf{X}$ and $\mathbf{Y}$ are real. In fact we can consider $\mathbf{X}$ and $\mathbf{Y}$ to be with complex or even quaternion entries where the norm of each entry has unit variance, $\mathbb{E}[x_i^t (y_j^t)^*]=c \delta_{i,j}$ and define $F=(\mathbf{X}\mathbf{Y}^\dagger+\mathbf{Y}\mathbf{X}^\dagger$)/(2T). The eigenvalue spectrum in the complex ($\beta=2$) and quaternion ($\beta=4$) cases is then the same as the real case. These are three cases of the same beta-ensemble where as usual the density is independent of beta but not the eigenvalue fluctuations and correlations. The matrix potential of this ensemble satisfies $V'(\lambda)=2\Re G_\mathbf{F}(\lambda)$ for $|\lambda|\leq\lambda_+$ where $G_\mathbf{F}(z)$ is the correct root of the cubic equation \eqref{cubic}, see e.g. \cite{CUP}.

Using $S$-transforms to deal with free products of matrices, we know that the Mar\v cenko-Pastur result can be generalised to the case where the ``true'' (population) covariance matrix of the $x$'s and of the $y$'s is a general definite positive matrix $\mathbf{C}$, i.e.:
\begin{equation}
        \widehat {\mathbf{X}} = \sqrt{\mathbf{C}} \mathbf{X}; \qquad \widehat {\mathbf{Y}} = \sqrt{\mathbf{C}} \mathbf{Y},
\end{equation}
with $N^{-1} \tr \mathbf{C} =1$. 

The same trick can be used in the present case as well, with now
\begin{equation}
\widehat{\mathbf{F}} = {\frac{1}{2T}} \left( \widehat{\mathbf{X}} \widehat{\mathbf{Y}}^\top + \widehat{\mathbf{Y}} \widehat {\mathbf{X}}^\top \right) = \frac{1}{2T} \sqrt{\mathbf{C}}\left( \mathbf{X} \mathbf{Y}^\top + \mathbf{Y} \mathbf{X}^\top \right) \sqrt{\mathbf{C}}.
\end{equation}
Interestingly, when $c=0$, one finds that the spectrum of $\widehat {\mathbf{F}}$ is the same as that of  $\widehat {\mathbf{F}}$ and therefore {\it independent} of $\mathbf{C}$. Indeed for a trace-less matrix (such as $\mathbf{F}$ with $c=0$) the free product with matrix $\mathbf{C}$ is equivalent to a simple scaling of $\mathbf{F}$ by $N^{-1} \tr \mathbf{C}=1$. For $c > 0$, however, this is not true -- as it is well known in the Mar\v cenko-Pastur case, see e.g. \cite{Bai,CUP}.

In conclusion, we have defined a new class of random correlation matrices with non-positive eigenvalues. We have determined the eigenvalue spectrum, which defines a two-parameter family of distributions that contain the Mar\v cenko-Pastur law in one limit ($c=1$, $q$ arbitrary) and the Wigner semi-circle in another limit ($c=0$, $q \to 0$). The $c=0$ case provides a null-hypothesis to test the existence of cross-correlations between two time series, complementing the results of \cite{Micelli,Florent}.

\vskip 0.3cm
We thank F. Benaych-Georges, M. Nowak and D. Savin for useful comments, in particular pointing us to references \cite{Nica,tetilla}.


\begin{thebibliography}{}
\bibitem{MP} V. A. Mar\v cenko and L. A. Pastur. Distribution of eigenvalues for some sets of
random matrices. Matematicheskii Sbornik, 114(4):507–536, 1967.

\bibitem{Wachter} K. W. Wachter. The limiting empirical measure of multiple discriminant ratios. The
Annals of Statistics, 8(5):937–957, 1980.

\bibitem{Micelli} J.-P. Bouchaud, L. Laloux, M. A. Miceli, and M. Potters. Large dimension forecasting
models and random singular value spectra. The European Physical Journal B,
55(2):201–207, 2007.

\bibitem{Verdu} A. M. Tulino and S. Verd\`u. Random Matrix Theory and Wireless Communications.
Now publishers, Hanover, Mass., 2004.

\bibitem{CUP} M. Potters, J. P. Bouchaud. A First Course In Random Matrix Theory, Cambridge University Press, in press, 2020.

\bibitem{Nica} A. Nica, R. Speicher, Commutators of free random variables. Duke Math. J. 92, no. 3, 553--592, 1998.

\bibitem{tetilla} A. Deya and  I. Nourdin. Convergence of Wigner integrals to the tetilla law.  Latin American Journal of Probability and Mathematical Statistics, Instituto Nacional de Matematica Pura e Aplicada, 9, pp.101-127, 2012.

\bibitem{Bai} J. W. Silverstein and Z. Bai. On the empirical distribution of eigenvalues of a
class of large dimensional random matrices. Journal of Multivariate Analysis,
54(2):175–192, 1995.

\bibitem{Florent} F. Benaych-Georges, J. P. Bouchaud and M. Potters. Optimal cleaning for singular values of cross-covariance matrices. preprint, arXiv:1901.05543, 2019.



\end{thebibliography}
\end{document}